\documentclass[conference,9pt]{IEEEtran}
\IEEEoverridecommandlockouts

\usepackage{cite}
\usepackage{amsmath,amssymb,amsfonts}
\usepackage{algorithmic}
\usepackage{booktabs}
\usepackage{makecell}
\usepackage{multirow}
\usepackage{tabularx}
\usepackage{array}
\usepackage{makecell}
\usepackage{graphicx}
\usepackage{textcomp}
\usepackage{xcolor}
\def\BibTeX{{\rm B\kern-.05em{\sc i\kern-.025em b}\kern-.08em
    T\kern-.1667em\lower.7ex\hbox{E}\kern-.125emX}}
\begin{document}

\title{Enhancing Low-Resource ASR through Versatile TTS: Bridging the Data Gap
}

\author{\IEEEauthorblockN{
Guanrou Yang$^{1}$, Fan Yu$^{2}$, Ziyang Ma$^{1}$, Zhihao Du$^{2}$, Zhifu Gao$^{2}$, Shiliang Zhang$^{2}$, Xie Chen$^{1 \dag}$\thanks{$\dag$ Corresponding Author}}

\IEEEauthorblockA{\textit{  $^1$MoE Key Lab of Artificial Intelligence, AI Institute, X-LANCE Lab,  Shanghai Jiao Tong University, China} \\
\textit{$^{2}$Institute for Intelligent Computing, Alibaba Group,  China}\\
\textit{ \{yangguanrou, chenxie95\}@sjtu.edu.cn}}
}

\maketitle

\begin{abstract}
While automatic speech recognition (ASR) systems have achieved remarkable performance with large-scale datasets, their efficacy remains inadequate in low-resource settings, encompassing dialects, accents, minority languages, and long-tail hotwords, domains with significant practical relevance. 
With the advent of versatile and powerful text-to-speech (TTS) models, capable of generating speech with human-level naturalness, expressiveness, and diverse speaker profiles, leveraging TTS for ASR data augmentation provides a cost-effective and practical approach to enhancing ASR performance. 
Comprehensive experiments on an unprecedentedly rich variety of low-resource datasets demonstrate consistent and substantial performance improvements, proving that the proposed method of enhancing low-resource ASR through a versatile TTS model is highly effective and has broad application prospects.  
Furthermore, we delve deeper into key characteristics of synthesized speech data that contribute to ASR improvement, examining factors such as text diversity, speaker diversity, and the volume of synthesized
data, with text diversity being studied for the first time in this work.
We hope our findings provide helpful guidance and reference for the practical application of TTS-based data augmentation and push the advancement of low-resource ASR one step further.


\end{abstract}


\section{Introduction}
Although ASR technology has seen remarkable advancements, achieving near-perfect word error rates (WER) on high-quality, large-scale datasets, ASR still faces significant challenges when applied to low-resource datasets~\cite{li2022recent}. These datasets encompass accented speech, regional languages or dialects, minority languages, and long-tail hotwords, which are of critical practical value.

Improving ASR models on low-resource datasets has been extensively researched. One widely used approach is self-training, where an ASR model, trained with limited human-transcribed data, is utilized to generate transcriptions, which are then used to train the ASR system further together with original annotated data~\cite{kahn2020self,chen2020semi}. While this method can enhance ASR performance, it has several limitations. For certain low-resource languages, the scarcity of textual data, let alone speech resources, renders the use of unlabeled speech for pseudo-labeling more of an idealized assumption than a practical solution. Besides, the model employed for labeling has to exhibit ideal performance to produce high-quality labels, which necessitates training with substantial data, making this approach somewhat circular. Furthermore, self-training typically requires multiple iterations to refine the ASR model and label quality continuously, leading to increased costs and extended experimental cycles. Moreover, the iterative self-training approach may prove less efficient, with minor performance gains compared with alternative methods, such as augmenting training data with TTS systems~\cite{bartelds2023making}.

Another promising approach is utilizing TTS systems to generate synthetic speech for data augmentation~\cite{laptev2020you}. This method stands out for its cost-effectiveness and practicality, requiring only textual data, which is generally more accessible than speech data. Even in cases of limited text availability, large language models or grammar-based techniques can be harnessed to generate it.
However, training an ASR model solely on synthetic speech has not been successful before for synthetic data were not realistic and representative enough~\cite{le2024voicebox}. Traditional TTS models~\cite{wang2017tacotron,kim2020glow,popov2021grad} commonly train on only dozens to hundreds of hours of single-speaker or multi-speaker data, producing speech with constrained timbre, prosody, and speaking styles. Input text and output speech often have rigid one-to-one mapping without randomness~\cite {wang2023neural}. Besides, previous TTS systems tend to produce clean, high-fidelity speech for better listening experiences, ill-suited for ASR training that benefits from more varied and noisy speech with diverse acoustic environments.

The rapid advancement of TTS technology and the recent emergence of a new generation of powerful models~\cite{du2024cosyvoice,anastassiou2024seed,lajszczak2024base,zhang2023speak,chen2024vall} have revolutionized ASR data augmentation. Trained on vast and diverse datasets, these models can generate speech with human-level naturalness and expressiveness, offering superior control over various attributes such as emotion, timbre, and style. This enables the production of highly expressive and diverse speech, adaptable to a wide range of speakers and scenarios. Consequently, leveraging TTS-generated data can effectively augment ASR training, providing a practical solution for addressing low-resource challenges in ASR system development. Recent experiments have shown that ASR models trained exclusively on synthetic speech generated by top-tier TTS models Seed-TTS~\cite{anastassiou2024seed} and Voicebox~\cite{le2024voicebox} can achieve competitive performance, with only minor gaps compared to models trained on real data, further confirming the potential for utilizing synthetic data in the advancement of speech understanding models.


In this paper, we revisit the use of versatile TTS models for data augmentation to enhance low-resource ASR systems comprehensively, claiming that the time has come for these models to play a pivotal role in advancing ASR performance in such scenarios.
Specifically, we employ a limited amount of ASR training data to fine-tune the pre-trained multilingual CosyVoice-base model and mix the synthesized speech with the original speech to augment the ASR training process. We conduct extensive experiments across various types of low-resource datasets, including accented English speech, minority languages including Korean, Chinese dialects containing Wu and Min Nan, and long-tail hotwords datasets in the automotive industry. The results consistently demonstrate significant performance improvements, validating the effectiveness of our approach. To gain deeper insight, we explore what characteristics of synthesized speech data play a critical role in ASR enhancement, focusing on factors such as text diversity, speaker diversity, and the quantity of synthesized data. The results show that while increasing TTS-augmented data consistently improves ASR performance, generating an amount of TTS data comparable to real data is cost-effective, as further production brings marginal improvement. Text diversity proves to have a more significant influence than speaker diversity, while a modest number of approximately 50 to 100 speakers is sufficient, challenging conventional expectations. The main contributions of this work are summarized as follows:
\begin{itemize}
\item  Unlike previous efforts limited to a single dataset or a specific type of dataset, we conduct comprehensive experiments across various low-resource datasets including dialects, accents, minority languages, and long-tail hotwords domains, demonstrating the generality and effectiveness of our approach.
\item We utilize a cutting-edge TTS model to synthesize speech with human-level naturalness, expressiveness, and diverse speaker profiles, resulting in significant ASR performance gains.
\item We systematically analyze the key factors influencing ultimate ASR performance—text diversity, speaker diversity, and data volume. Notably, our thorough investigation and analysis of the text and speaker diversity of synthesized speech is unprecedented in prior work. Counter-intuitive conclusions and practical guidance are provided for further TTS-augmented ASR systems development. 
\end{itemize}

\begin{figure*}[t]
  \centering
  \includegraphics[width=\linewidth]{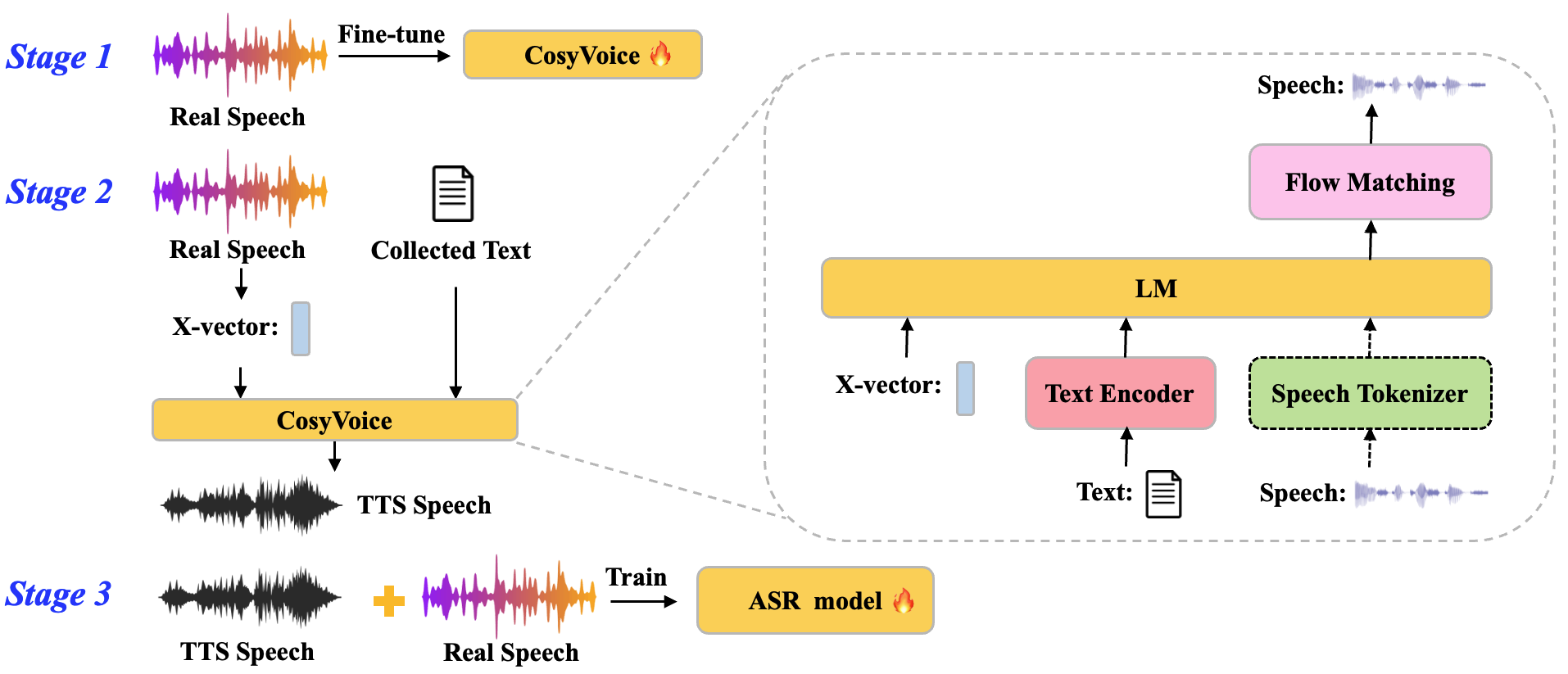}
  \caption{Pipeline of proposed TTS-augmented ASR training method and the architecture of CosyVoice model.}
  \label{fig:pipeline}
  \vspace{-5mm} 
\end{figure*}


\vspace{-2mm} 
\section{Related Works}
Previous research has explored TTS data augmentation in various domains, such as ASR for children’s speech~\cite{rolland2024improved}, conversational speech~\cite{cornell2024generating}, minority language speech~\cite{gokay2019improving}, dysarthic speech~\cite{leung2024training} and other tasks such as speech emotion recognition~\cite{ma2024leveraging,malik2023preliminary} and keyword spotting~\cite{park2024utilizing}.

Beyond demonstrating the effectiveness of this approach across different speech tasks and diverse domains, a series of works delve deeply into more efficient ways of leveraging the imperfect TTS data, addressing the distribution gaps and domain mismatches between synthetic and real data. For instance, Ma et al.~\cite{ma2024leveraging} introduce four distinct methods for integrating TTS data with real data including curriculum learning and transfer learning. Rolland et al.~\cite{rolland2024improved} validate the effectiveness of adapter-based architectures in using imperfect TTS data. Hu et al.~\cite{hu2022synt++} propose a rejection sampling algorithm and utilize separate batch normalization statistics for real and TTS data. Liu et al.~\cite{liu2023towards} select synthetic speech dissimilar to real speech via a neural network, effectively boosting recognition performance.

Since our approach of fine-tuning CosyVoice and a straightforward combination of real and TTS data achieves superior results without employing complex strategies, we focus on a different question: which characteristics of TTS data are most beneficial for augmenting ASR training? Several existing studies have provided preliminary answers and insights.~\cite{rossenbach2024problem} discover that the NISQA MOS~\cite{mittag2021deep} and synthetic WER have no clear relations to the ASR performance, aligning with observations in~\cite{anastassiou2024seed} that lower WER typically indicates more standard speech easier for recognition but lose distinctive qualities like strong accent or high expressiveness and may not be helpful for ultimate ASR training. ~\cite{du2020speaker} proposes a speaker augmentation approach for TTS data augmentation to improve speaker diversity of synthetic speech. Experiments in ~\cite{du2024cosyvoice} reveal that text diversity is more critical than the duration of speech for the ASR task. Inspired by these related works, we posit the diversity of synthesized speech is a pivotal factor influencing the ultimate ASR performance and endeavor to investigate the impacts of text diversity and speaker diversity on the efficacy of TTS data augmentation, alongside the influence of the quantity of synthetic data employed.

\section{Methods}
\subsection{Overview of CosyVoice}
CosyVoice is a scalable multilingual zero-shot TTS system that builds upon supervised semantic tokens derived from a multilingual ASR model. Unlike conventional unsupervised methods, which lack explicit alignment between text and speech, CosyVoice employs supervised tokens to enhance content consistency and speaker similarity. As shown in Figure \ref{fig:pipeline}, the model architecture comprises four key components: a text encoder, a speech tokenizer, a large language model (LLM), and a conditional flow matching model. The text encoder ensures alignment between text and speech tokens, while the speech tokenizer extracts semantic tokens from speech input using vector quantization techniques. The core of the model is an LLM, which models TTS as an auto-regressive sequence generation task, predicting speech tokens based on text prompts. Finally, a conditional flow matching model synthesizes speech from the generated tokens.

In addition to the enhanced content consistency and speaker similarity through the use of supervised semantic tokens, CosyVoice leverages x-vectors~\cite{snyder2018x} to disentangle the multifaceted aspects of speech, including semantic content, speaker identity, and prosody. This enables the model to capture subtle details like speaker timbre and environmental factors, producing more natural and diverse speech. Consequently, CosyVoice is capable of synthesizing speech with a wide range of voices and speaking styles, a feature critical for developing robust ASR systems.


\subsection{Proposed Pipeline}
We propose a succinct three-stage pipeline for leveraging TTS-generated data to augment ASR training datasets, as illustrated in Figure \ref{fig:pipeline}. In the first stage, we fine-tune the CosyVoice model with limited ASR training data, specifically adjusting its language model and conditional flow matching model separately. The fine-tuning step tailors the pre-trained CosyVoice model to better capture the acoustic and linguistic characteristics of the low-resource target data of real-world scenarios. In the second stage, for each text input, we generate synthetic speech by applying a different speaker vector, effectively creating diverse utterances with various speakers. In the third stage, this enriched synthetic data is integrated with the original training data to create a more comprehensive and diverse dataset for follow-up ASR training. We employ the standard Conformer~\cite{gulati2020conformer} ASR model to evaluate the quality and effectiveness of the generated speech.


\subsection{Model Configuration and Training}


We utilize the open-source CosyVoice-base-300M model as our pre-trained model, which is trained on a large-scale multi-lingual dataset comprising 130k hours of Mandarin, 30k hours of English, 5k hours of Cantonese, 4.6k hours of Japanese, and 2.2k hours of Korean content.
The two integral components of CosyVoice, the LLM and the conditional Flow Matching Model, can be fine-tuned separately. 

The text encoder is composed of 8 Transformer layers. The LLM has a similar structure but a deeper architecture, consisting of 16 Transformer layers and an additional linear output decoder layer. The text encoder and the LLM have a total of 343.20 million parameters. 

The speech encoder includes 6 Transformer layers, while the conditional flow matching model primarily comprises ResNet blocks and Transformer blocks. The decoder also includes Conv1D layers as downsampling and upsampling layers to adjust the resolution of the input features. The combined speech encoder and conditional flow matching model consist of 104.87 million parameters.
The speech tokenizer is built on the proprietary SenseVoice-Large ASR model~\cite{speechteam2024funaudiollm,radford2023robust}, which places a vector quantizer after the first six encoder layers, using a single 4,096-entry codebook. 

For fine-tuning these two modules, we use AdamW as the optimizer with betas set to (0.9, 0.999) and a weight decay of 0.01. The learning rate is initially set to 1e-4 and gradually increases during the warmup phase with 10,000 warmup steps.

During the inference procedure of CosyVoice, we employ the top-k sampling strategy for decoding and set the k as 5 to introduce certain diversity in the output.

For the ASR training procedure, we employ a standard ASR architecture consisting of a 12-block Conformer encoder, a 6-block Transformer decoder, and a hybrid CTC/attention loss, with approximately 120 million parameters in total. We extract fbank features from input waveforms and apply specaugmentation~\cite{park2019specaugment}. We use Adam as the optimizer and the learning rate increases linearly from 0 to 5e-4 in the initial 25k steps and then decays exponentially to half of the peak rate in the remaining steps.

\section{Experiments}
\subsection{Experimental Setup}
\subsubsection{Dataset}
For the accented English data, we employ the Common Voice English dataset\cite{ardila2019common}. We mainly utilize the eight high-resource accents in this dataset, including Australia, Canada, England, India, Ireland, New Zealand, Scotland, and the United States. The training set split follows the same procedure as \cite{lin2023wav2vec}.

The remaining datasets are in-house annotated data within the industry.
For Chinese dialects, the Wu dataset is collected mainly from Suzhou, Hangzhou, and Shanghai. About 600 speakers participate in the recording process, consisting of equal numbers of male and female speakers. 
The recordings are conducted in both quiet and noisy environments to ensure robustness across different acoustic conditions. The recorded texts encompass a variety of everyday conversational topics, readings from news, and social media texts such as tweets, providing diverse linguistic contexts. Similarly, the Min Nan dialect dataset is compiled with a comparable methodology.
The hotwords dataset is collected from real-world conversational data related to the automotive industry. It features a wealth of hotwords, specifically automotive terminology encompassing vehicle models and configurations, alongside sales and financial terms. The Korean dataset is collected from multiple domains, including literature, education, news, sports etc.

\begin{table}[h!]
\centering
\caption{ASR Performance with TTS Data Augmentation across different Data Types and Languages.}
\begin{tabular}{llcccc}
\toprule
\textbf{Data Type} & \textbf{Language} & \textbf{Real(h)} & \textbf{TTS(h)} & \textbf{\makecell{WER\\ /CER(\%)}} & \textbf{\makecell{ Rel \\ Gain(\%)}} \\ 
\midrule
\multirow{2}{*}{\makecell{Accented\\English}} & \multirow{2}{*}{English} & 300 & 0 & 12.61 & \multirow{2}{*}{15.23} \\
                                  &  & 300 & 300 & 10.69 & \\
\midrule
   \multirow{2}{*}{\makecell{Minority\\Language}} & \multirow{2}{*}{Korean} & 3700 & 0 & 8.48 & \multirow{2}{*}{25.59} \\
                                  &  & 3700 & 4000 & 6.31 & \\
\midrule
\multirow{4}{*}{\makecell{Chinese\\Dialects}} & \multirow{2}{*}{Min Nan} & 200 & 0 & 55.42 & \multirow{2}{*}{45.85} \\
                                  &  & 200 & 200 & 30.00 & \\
\cmidrule{2-6}
                                  & \multirow{2}{*}{Wu} & 200 & 0 & 27.47 & \multirow{2}{*}{27.45} \\
                                  &  & 200 & 200 & 19.93 & \\
\midrule
\multirow{2}{*}{Hotwords} & \multirow{2}{*}{Chinese} & 300 & 0 & 21.32 & \multirow{2}{*}{38.33} \\
                          &  & 300 & 300 & 13.15 & \\
\bottomrule
\end{tabular}
\vspace{-5mm} 
\label{tab:all}
\end{table}

\subsection{Experimental Results}
\vspace{-1mm} 
To evaluate the impact of TTS-generated data augmentation on ASR training, we conduct experiments across a variety of low-resource datasets, including different languages and dialects. The results are summarized in Table \ref{tab:all}.
For accented English, we use a dataset containing 300 hours of real English speech. Without TTS augmentation, the ASR system achieves a WER of 12.61. When 300 additional hours of TTS-generated data are incorporated, the WER is reduced to 10.69. 
For minority languages, we perform experiments on Korean. 
Similarly, for Korean, the CER decreases from 8.48 to 6.31 when 4,000 hours of TTS-generated speech is added to the original 3,700 hours of real speech. In the case of Chinese dialects, we experiment with Min Nan and Wu. For Min Nan, using 200 hours of real speech results in a high CER of 55.42. Augmenting the training set with an additional 200 hours of TTS-generated data leads to a substantial WER reduction to 30.00. For Wu, a similar pattern is observed, with the CER dropping from 27.47 to 19.93 when TTS data is included.
Lastly, for a Chinese hotword dataset with 300 hours of real speech, the ASR system initially achieves a CER of 21.32. After augmenting the data with an additional 300 hours of TTS-generated speech, the CER is reduced to 13.15.

The results demonstrate that TTS-generated data can effectively enhance ASR training on low-resource languages and dialects. While the recognition difficulty varies across datasets and the extent of WER/CER improvement differs, we observe consistent improvements in WER/CER across all datasets when TTS data is incorporated into the training. These findings highlight the potential of TTS data augmentation to improve ASR systems, especially in challenging low-resource scenarios.

\vspace{-3mm} 
\begin{table}[h!]
\centering
\caption{ASR performance with different amounts of TTS augmented data on the Korean dataset.}
\begin{tabular}{lcc}
\toprule
\textbf{Real (h)} & \textbf{TTS (h)} &\textbf{ WER (\%)} \\ 
\midrule
\multirow{6}{*}{3700} &0 & 8.48 \\
                      & 1000 & 7.57 \\
                      & 2000 & 6.75 \\
                      & 3000 & 6.51 \\
                      & 4000 & 6.31 \\
                      & 5000 & 6.18 \\
\bottomrule
\end{tabular}
\vspace{-3mm} 
\label{tab:amount}
\end{table}

\vspace{-2mm} 
\subsection{Ablation Study}
To delve deeper into which characteristics of synthesized speech play a role in enhancing ASR training, we specifically examine the influence of the quantity, speaker diversity, and text diversity of generated speech, on the ultimate ASR performance. The experimental results are respectively presented in Tables \ref{tab:amount}, \ref{tab:speaker}, \ref{tab:text}.

To explore the impact of augmenting ASR training with different amounts of TTS-generated data, we synthesize varying durations of Korean speech. 
The experimental results from table \ref{tab:amount} demonstrate that increasing the amount of TTS-augmented data consistently improves ASR performance, as indicated by the decreasing CER. The greatest improvement occurs when comparing the baseline with no TTS augmentation (8.48 CER) to the model trained with 5000 hours of TTS data (6.18 CER). However, the performance gains begin to converge as the amount of TTS data increases, indicating that the benefits may taper off after a certain point. Consequently, generating an amount of TTS data equivalent to real data may suffice, as additional production incurs higher costs with marginal improvement.


\begin{table}[htbp]
\vspace{-3mm} 
\centering
\caption{ASR Performance with TTS Data Augmentation at different levels of Speaker Diversity on Min Nan Dataset. The same texts are used to generate an equal amount of speech, with only the number of speakers varying. Mean Pairwise Cosine Sim. with * denotes computing after K-means clustering.}
\label{tab:speaker}
\begin{tabular}{ccc}
\toprule
\textbf{\#Speaker for TTS} & \textbf{Mean Pairwise Cosine Sim.} & \textbf{CER (\%)} \\
\midrule
w/o TTS Augmentation & N/A & 55.42 \\
\midrule
200k & 0.1676* & 30.00 \\
100k & 0.1821* & 29.31 \\
50k & 0.2028* & 27.98 \\
20k & 0.2384* & 28.34 \\
10k & 0.2727* & 27.98 \\
1k & 0.1213 & 27.81 \\
100 & 0.1212 & 28.11 \\
50 & 0.1228 & 29.64 \\
10 & 0.1950 & 36.61 \\
5 & 0.2436 & 32.81 \\
1 & N/A & 39.18 \\
\bottomrule
\end{tabular}
\vspace{-3mm} 
\end{table}

To investigate the role of speaker diversity, 
we keep the total amount of text used for speech synthesis constant, varying only the number of distinct speaker vectors utilized. Specifically, we randomly sample different proportions of x-vectors from 200k x-vectors extracted from the original 200k real speech of the Min Nan dataset. The mean cosine similarity between all pairs of speaker vectors is calculated to quantify the diversity of the employed speaker vectors. For experiments involving more than 1,000 speaker vectors, we first employ K-means clustering, setting the number of clusters to 5\% of the total speaker vectors. Then, we compute the Mean Pairwise Cosine Similarity on the cluster centroids. 

From the experimental results in Table \ref{tab:speaker}, we reach an intriguing counter-intuitive conclusion: we do not require as many different speaker vectors as expected. A diversity of around 50 to 100 distinct speakers is sufficient. Employing a limited number of speaker vectors but generating multiple utterances for each appears to yield better results than using all available speaker vectors with only one speech per speaker.
Besides, as the number of speakers increases, the Mean Pairwise Cosine Similarity decreases, indicating higher diversity among the synthesized voices. An insufficient number of speaker vectors yields unsatisfactory performance, consistent with previous findings where traditional TTS models with single or few speakers are ineffective for ASR data augmentation.

\begin{table}[htbp]
\vspace{-3mm} 
\centering
\caption{ASR Performance with TTS Data Augmentation at different levels of Text Diversity on the Min Nan Dataset. The number of speakers and the amount of generated speech remain constant, with only the texts used varying.}
\label{tab:text}
\begin{tabular}{ccccc}
\toprule
\textbf{\#Sentence for TTS} & \textbf{TTR (\%)} & \textbf{Entropy} & \textbf{PPL} & \textbf{CER (\%)} \\
\midrule
w/o TTS Augmentation & N/A &  N/A& N/A&55.42 \\
\midrule
200k & 0.189 & 9.522 & 10.00 & 30.00 \\
100k & 0.178 & 9.520 & 10.30 & 32.45 \\
50k & 0.164 & 9.513 & 10.31 & 30.35 \\
20k & 0.142 & 9.507 & 10.38 & 35.33 \\
10k & 0.125 & 9.486 & 10.26 & 38.57 \\
5k & 0.108 & 9.481 & 10.31 & 48.89 \\
1k & 0.067 & 9.361 & 10.06 & 55.14 \\
500 & 0.052 & 9.243 & 9.58 & 52.77 \\
\bottomrule
\end{tabular}
\vspace{-3mm} 
\end{table}

To examine the effect of text diversity, we randomly select varying proportions of sentences from the whole Min Nan dataset. We use Type-Token Ratio (TTR), Entropy, and Perplexity (PPL) as metrics to evaluate text diversity. The results in Table \ref{tab:text} suggest that text diversity exerts a more substantial impact than speaker diversity, with ASR performance more sensitive to variations in text diversity. Specifically, while moderate reductions in text diversity (down to 50,000 sentences) still yield reasonable ASR accuracy, further reductions (below 10,000 sentences) lead to a marked increase in CER. Therefore, maintaining a sufficient level of text diversity is essential for augmenting ASR training.

\vspace{-1mm} 
\section{Conclusion}
\vspace{-1mm} 
In this work, we emphasize that the rapid advancements in TTS technology, especially the rise of large-scale, versatile models, have opened up unprecedented opportunities for enhancing ASR systems in low-resource settings. The improved synthesis quality, diversity, and controllability of modern TTS models have transformed the landscape, offering new avenues for addressing the persistent challenges of lacking training data in ASR research. 
Through extensive experiments, we observe substantial performance gains across various datasets by generating diverse and high-quality synthetic speech with CosyVoice. Key factors such as text diversity, speaker diversity, and the volume of synthesized data are systematically analyzed, revealing that text diversity plays a more critical role than speaker diversity in enhancing ASR performance, and generating a moderate amount of TTS data, comparable to the original data provides the highest cost-efficiency. These findings underscore the potential of advanced TTS models as a practical and efficient solution to the persistent challenges of low-resource ASR development and provide helpful insights for future research and real-world applications.

\bibliographystyle{IEEEtran}
\bibliography{refs}

\begin{thebibliography}{10}
\providecommand{\url}[1]{#1}
\csname url@samestyle\endcsname
\providecommand{\newblock}{\relax}
\providecommand{\bibinfo}[2]{#2}
\providecommand{\BIBentrySTDinterwordspacing}{\spaceskip=0pt\relax}
\providecommand{\BIBentryALTinterwordstretchfactor}{4}
\providecommand{\BIBentryALTinterwordspacing}{\spaceskip=\fontdimen2\font plus
\BIBentryALTinterwordstretchfactor\fontdimen3\font minus \fontdimen4\font\relax}
\providecommand{\BIBforeignlanguage}[2]{{%
\expandafter\ifx\csname l@#1\endcsname\relax
\typeout{** WARNING: IEEEtran.bst: No hyphenation pattern has been}%
\typeout{** loaded for the language `#1'. Using the pattern for}%
\typeout{** the default language instead.}%
\else
\language=\csname l@#1\endcsname
\fi
#2}}
\providecommand{\BIBdecl}{\relax}
\BIBdecl

\bibitem{li2022recent}
J.~Li \emph{et~al.}, ``Recent advances in end-to-end automatic speech recognition,'' \emph{Proc. APSIPA}, 2022.

\bibitem{kahn2020self}
J.~Kahn, A.~Lee, and A.~Hannun, ``Self-training for end-to-end speech recognition,'' in \emph{Proc. ICASSP}, 2020.

\bibitem{chen2020semi}
Y.~Chen, W.~Wang, and C.~Wang, ``Semi-supervised {ASR} by end-to-end self-training,'' \emph{arXiv preprint}, 2020.

\bibitem{bartelds2023making}
M.~Bartelds, N.~San, B.~McDonnell, D.~Jurafsky, and M.~Wieling, ``Making more of little data: Improving low-resource automatic speech recognition using data augmentation,'' \emph{arXiv preprint}, 2023.

\bibitem{laptev2020you}
A.~Laptev, R.~Korostik, A.~Svischev, A.~Andrusenko, I.~Medennikov, and S.~Rybin, ``You do not need more data: Improving end-to-end speech recognition by text-to-speech data augmentation,'' in \emph{Proc. CISP-BMEI}, 2020.

\bibitem{le2024voicebox}
M.~Le, A.~Vyas, B.~Shi, B.~Karrer, L.~Sari, R.~Moritz, M.~Williamson, V.~Manohar, Y.~Adi, J.~Mahadeokar \emph{et~al.}, ``Voicebox: Text-guided multilingual universal speech generation at scale,'' \emph{Advances in neural information processing systems}, 2024.

\bibitem{wang2017tacotron}
Y.~Wang, R.~Skerry-Ryan, D.~Stanton, Y.~Wu, R.~J. Weiss, N.~Jaitly, Z.~Yang, Y.~Xiao, Z.~Chen, S.~Bengio \emph{et~al.}, ``Tacotron: Towards end-to-end speech synthesis,'' \emph{arXiv preprint}, 2017.

\bibitem{kim2020glow}
J.~Kim, S.~Kim, J.~Kong, and S.~Yoon, ``Glow-tts: A generative flow for text-to-speech via monotonic alignment search,'' \emph{Proc. NeurIPS}, 2020.

\bibitem{popov2021grad}
V.~Popov, I.~Vovk, V.~Gogoryan, T.~Sadekova, and M.~Kudinov, ``Grad-tts: A diffusion probabilistic model for text-to-speech,'' in \emph{Proc. ICML}, 2021.

\bibitem{wang2023neural}
C.~Wang, S.~Chen, Y.~Wu, Z.~Zhang, L.~Zhou, S.~Liu, Z.~Chen, Y.~Liu, H.~Wang, J.~Li \emph{et~al.}, ``Neural codec language models are zero-shot text to speech synthesizers,'' \emph{arXiv preprint}, 2023.

\bibitem{du2024cosyvoice}
Z.~Du, Q.~Chen, S.~Zhang, K.~Hu, H.~Lu, Y.~Yang, H.~Hu, S.~Zheng, Y.~Gu, Z.~Ma \emph{et~al.}, ``Cosyvoice: A scalable multilingual zero-shot text-to-speech synthesizer based on supervised semantic tokens,'' \emph{arXiv preprint}, 2024.

\bibitem{anastassiou2024seed}
P.~Anastassiou, J.~Chen, J.~Chen, Y.~Chen, Z.~Chen, Z.~Chen, J.~Cong, L.~Deng, C.~Ding, L.~Gao \emph{et~al.}, ``{Seed-TTS}: A family of high-quality versatile speech generation models,'' \emph{arXiv preprint}, 2024.

\bibitem{lajszczak2024base}
M.~{\L}ajszczak, G.~C{\'a}mbara, Y.~Li, F.~Beyhan, A.~van Korlaar, F.~Yang, A.~Joly, {\'A}.~Mart{\'\i}n-Cortinas, A.~Abbas, A.~Michalski \emph{et~al.}, ``{BASE TTS}: Lessons from building a billion-parameter text-to-speech model on 100k hours of data,'' \emph{arXiv preprint}, 2024.

\bibitem{zhang2023speak}
Z.~Zhang, L.~Zhou, C.~Wang, S.~Chen, Y.~Wu, S.~Liu, Z.~Chen, Y.~Liu, H.~Wang, J.~Li \emph{et~al.}, ``Speak foreign languages with your own voice: Cross-lingual neural codec language modeling,'' \emph{arXiv preprint}, 2023.

\bibitem{chen2024vall}
S.~Chen, S.~Liu, L.~Zhou, Y.~Liu, X.~Tan, J.~Li, S.~Zhao, Y.~Qian, and F.~Wei, ``{VALL-E 2}: Neural codec language models are human parity zero-shot text to speech synthesizers,'' \emph{arXiv preprint}, 2024.

\bibitem{rolland2024improved}
T.~Rolland and A.~Abad, ``Improved children’s automatic speech recognition combining adapters and synthetic data augmentation,'' in \emph{Proc. ICASSP}, 2024.

\bibitem{cornell2024generating}
S.~Cornell, J.~Darefsky, Z.~Duan, and S.~Watanabe, ``Generating data with text-to-speech and large-language models for conversational speech recognition,'' \emph{arXiv preprint}, 2024.

\bibitem{gokay2019improving}
R.~Gokay and H.~Yalcin, ``Improving low resource turkish speech recognition with data augmentation and {TTS},'' in \emph{Proc. SSD}, 2019.

\bibitem{leung2024training}
W.-Z. Leung, M.~Cross, A.~Ragni, and S.~Goetze, ``Training data augmentation for dysarthric automatic speech recognition by text-to-dysarthric-speech synthesis,'' \emph{arXiv preprint}, 2024.

\bibitem{ma2024leveraging}
Z.~Ma, W.~Wu, Z.~Zheng, Y.~Guo, Q.~Chen, S.~Zhang, and X.~Chen, ``Leveraging speech ptm, text llm, and emotional tts for speech emotion recognition,'' in \emph{Proc. ICASSP}, 2024.

\bibitem{malik2023preliminary}
I.~Malik, S.~Latif, R.~Jurdak, and B.~Schuller, ``A preliminary study on augmenting speech emotion recognition using a diffusion model,'' \emph{arXiv preprint}, 2023.

\bibitem{park2024utilizing}
H.~J. Park, D.~Agarwal, N.~Chen, R.~Sun, K.~Partridge, J.~Chen, H.~Zhang, P.~Zhu, J.~Bartel, K.~Kastner \emph{et~al.}, ``Utilizing {TTS} synthesized data for efficient development of keyword spotting model,'' \emph{arXiv preprint}, 2024.

\bibitem{hu2022synt++}
T.-Y. Hu, M.~Armandpour, A.~Shrivastava, J.-H.~R. Chang, H.~Koppula, and O.~Tuzel, ``Synt++: Utilizing imperfect synthetic data to improve speech recognition,'' in \emph{Proc. ICASSP}, 2022.

\bibitem{liu2023towards}
S.~Liu, L.~Sar{\i}, C.~Wu, G.~Keren, Y.~Shangguan, J.~Mahadeokar, and O.~Kalinli, ``Towards selection of text-to-speech data to augment {ASR} training,'' \emph{arXiv preprint}, 2023.

\bibitem{rossenbach2024problem}
N.~Rossenbach, R.~Schl{\"u}ter, and S.~Sakti, ``On the problem of text-to-speech model selection for synthetic data generation in automatic speech recognition,'' \emph{arXiv preprint}, 2024.

\bibitem{mittag2021deep}
G.~Mittag and S.~M{\"o}ller, ``Deep learning based assessment of synthetic speech naturalness,'' \emph{arXiv preprint}, 2021.

\bibitem{du2020speaker}
C.~Du and K.~Yu, ``Speaker augmentation for low resource speech recognition,'' in \emph{Proc. ICASSP}, 2020.

\bibitem{snyder2018x}
D.~Snyder, D.~Garcia-Romero, G.~Sell, D.~Povey, and S.~Khudanpur, ``X-vectors: Robust dnn embeddings for speaker recognition,'' in \emph{Proc. ICASSP}, 2018.

\bibitem{gulati2020conformer}
A.~Gulati, J.~Qin, C.-C. Chiu, N.~Parmar, Y.~Zhang, J.~Yu, W.~Han, S.~Wang, Z.~Zhang, Y.~Wu \emph{et~al.}, ``Conformer: Convolution-augmented transformer for speech recognition,'' \emph{arXiv preprint arXiv:2005.08100}, 2020.

\bibitem{speechteam2024funaudiollm}
T.~SpeechTeam, ``{FunAudioLLM: Voice Understanding and Generation Foundation Models for Natural Interaction Between Humans and LLMs},'' \emph{arXiv preprint}, 2024.

\bibitem{radford2023robust}
A.~Radford, J.~W. Kim, T.~Xu, G.~Brockman, C.~McLeavey, and I.~Sutskever, ``Robust speech recognition via large-scale weak supervision,'' in \emph{Proc. ICML}, 2023.

\bibitem{park2019specaugment}
D.~S. Park, W.~Chan, Y.~Zhang, C.-C. Chiu, B.~Zoph, E.~D. Cubuk, and Q.~V. Le, ``{SpecAugment}: A simple data augmentation method for automatic speech recognition,'' in \emph{Proc. Interspeech}, 2019.

\bibitem{ardila2019common}
R.~Ardila, M.~Branson, K.~Davis, M.~Henretty, M.~Kohler, J.~Meyer, R.~Morais, L.~Saunders, F.~M. Tyers, and G.~Weber, ``Common voice: A massively-multilingual speech corpus,'' \emph{arXiv preprint}, 2019.

\bibitem{lin2023wav2vec}
Y.~Lin, S.~Zhang, Z.~Gao, L.~Wang, Y.~Yang, and J.~Dang, ``Wav2vec-moe: An unsupervised pre-training and adaptation method for multi-accent {ASR},'' \emph{Electronics Letters}, 2023.

\end{thebibliography}

\end{document}